\documentclass{article}

\usepackage{arxiv}

\usepackage[utf8]{inputenc} 
\usepackage[T1]{fontenc}    
\usepackage{hyperref}       
\usepackage{url}            
\usepackage{booktabs}       
\usepackage{amsfonts}       
\usepackage{nicefrac}       
\usepackage{microtype}      
\usepackage{graphicx}
\usepackage[numbers]{natbib}
\usepackage{doi}

\usepackage{float}
\usepackage{hyperref} 
\usepackage{amsmath} 

\usepackage{algorithm} 
\usepackage{algpseudocode}

\title{CipherFace: A Fully Homomorphic Encryption-Driven Framework for Secure Cloud-Based Facial Recognition}

\author{
    \href{https://orcid.org/0000-0002-0345-0088}{\includegraphics[scale=0.06]{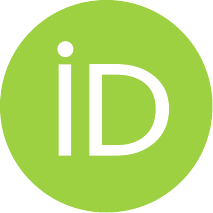}\hspace{1mm}Sefik Serengil} \\
	Solution Engineering\\
	Vorboss Limited\\
	London, UK \\
	\texttt{sefik.serengil@vorboss.com} \\
	\And
	\href{https://orcid.org/0000-0003-1250-5949}{\includegraphics[scale=0.06]{orcid.pdf}\hspace{1mm}Alper Ozpinar} \\
	Department of Business\\
	Ibn Haldun University\\
	Istanbul, Turkiye \\
	\texttt{alper@ozpinar.org} \\
}

\renewcommand{\shorttitle}{CipherFace}

\hypersetup{
pdftitle={CipherFace: A Fully Homomorphic Encryption-Driven Framework for Secure Cloud-Based Facial Recognition},
pdfsubject={cs.CR, cs.LG},
pdfauthor={Sefik Serengik, Alper Ozpinar},
pdfkeywords={Facial Recognition, Homomorphic Encryption},
}

\begin{document}
\maketitle

\begin{abstract}
Facial recognition systems rely on embeddings to represent facial images and determine identity by verifying if the distance between embeddings is below a pre-tuned threshold. While embeddings are not reversible to original images, they still contain sensitive information, making their security critical. Traditional encryption methods like AES are limited in securely utilizing cloud computational power for distance calculations. Homomorphic Encryption, allowing calculations on encrypted data, offers a robust alternative. This paper introduces CipherFace, a homomorphic encryption-driven framework for secure cloud-based facial recognition, which we have open-sourced at \url{http://github.com/serengil/cipherface}. By leveraging FHE, CipherFace ensures the privacy of embeddings while utilizing the cloud for efficient distance computation. Furthermore, we propose a novel encrypted distance computation method for both Euclidean and Cosine distances, addressing key challenges in performing secure similarity calculations on encrypted data. We also conducted experiments with different facial recognition models, various embedding sizes, and cryptosystem configurations, demonstrating the scalability and effectiveness of CipherFace in real-world applications.
\end{abstract}


\section{Introduction}

Facial recognition systems have become integral to various security and authentication applications, leveraging machine learning models to generate multi-dimensional embeddings that represent facial images \cite{vggface}. These embeddings are numerical representations of facial features that enable the identification and verification of individuals by comparing them against a stored database \cite{facenet}. The distance between embeddings, typically measured using metrics such as cosine distance or Euclidean distance \cite{qian2004similarity}, serves as the key criterion in distinguishing between different faces. For a pair of images of the same individual, their corresponding embeddings should exhibit a small distance, while embeddings of images from different individuals should have a large distance. By establishing a pre-tuned threshold, facial recognition systems can classify images as representing the same person or different individuals.

Although embeddings do not allow for the restoration of the original image, they still contain private and sensitive information akin to a person's fingerprint. As a result, facial embeddings can be vulnerable to attacks \cite{dong2019efficient}. If an adversary gains access to the pre-calculated embedding of an individual, they may potentially launch adversarial attacks, compromising the privacy and security of the system. Encrypting embeddings is a crucial step to mitigate such risks and ensure data security, particularly in cloud-based systems.

One option for encrypting embeddings is to use symmetric key algorithms, such as AES \cite{lu2009face}. However, this approach has practical limitations in terms of system management. The private key required for decryption must remain secure and should not be transmitted to the cloud system, as this could result in key leakage. If encrypted embeddings are pulled back to on-premises systems for decryption and distance calculation, the computation power of the cloud system is not utilized effectively.

Homomorphic encryption is an advanced cryptographic technique that allows computations to be performed directly on encrypted data without requiring decryption \cite{rivest1978data}. Unlike traditional encryption methods, where data must be decrypted before performing operations, homomorphic encryption ensures that sensitive information remains secure throughout the computation process. The results of these computations, when decrypted, are equivalent to those obtained if the operations were performed on unencrypted data.

Homomorphic encryption algorithms can be categorized as either partially or fully homomorphic. Partial homomorphic encryption \cite{phe} allows the system to perform only one type of operation—either addition or multiplication—on encrypted data. Many well-known public key algorithms are partially homomorphic. For example, RSA and ElGamal are multiplicatively homomorphic, while Exponential ElGamal, Elliptic Curve ElGamal, Paillier, Damgard-Jurik, Okamoto–Uchiyama, Benaloh, and Naccache–Stern are additively homomorphic \cite{lightphe}.

In contrast, fully homomorphic encryption \cite{fhe} enables both addition and multiplication operations on encrypted data. This capability makes fully homomorphic encryption significantly more versatile but also computationally intensive, as it requires much larger keys and greater computational power. Since distance calculations in facial recognition involve both addition and multiplication, fully homomorphic encryption is essential for securing embeddings while preserving the ability to compute distances in the encrypted domain.

Fully homomorphic encryption offers a more efficient and secure alternative for system design. In the context of facial recognition, a cloud system can calculate the encrypted distance between two encrypted embeddings without having access to the private key. The decrypted distance can then be evaluated on-premises to determine whether the two embeddings belong to the same person. This approach ensures both the security of embeddings and efficient utilization of the computational power of cloud systems.

In this paper, we propose CipherFace, a novel homomorphic encryption-driven framework aimed at securing facial embeddings in cloud-based facial recognition systems, open-sourced at \url{http://github.com/serengil/cipherface}. By leveraging the computational power of cloud systems while maintaining data privacy, CipherFace addresses critical challenges in securing biometric data. Our framework utilizes the DeepFace \cite{deepface2} library for Python to generate facial embeddings and the TenSEAL \cite{tenseal} library to perform fully homomorphic encryption, enabling secure storage and processing of facial recognition data.

A key contribution of this work is the development of novel encrypted distance computation schemes for both Euclidean and Cosine distances, overcoming inherent limitations in fully homomorphic encryption to enable secure and efficient comparison of encrypted embeddings. While CipherFace is specifically designed for facial recognition, its applicability extends to a wide range of similarity search use cases beyond biometrics. Potential applications include secure document comparison for plagiarism detection, privacy-preserving recommendation systems in e-commerce, encrypted medical data analysis for diagnostics, and fraud detection in financial transactions. The framework’s ability to process encrypted embeddings makes it suitable for any domain requiring secure and efficient similarity matching, thereby broadening its impact across multiple industries.

Furthermore, we conducted extensive experiments with different facial recognition models, various embedding sizes, and different cryptosystem configurations. These experiments demonstrate the effectiveness and scalability of CipherFace, making it a practical solution for real-world applications that require both robust performance and stringent data privacy guarantees.

\section{Related Work and Enhancements}

The research "Ciphertext Face Recognition System Based on Secure Inner Product Protocol" \cite{related2} combines Pallier homomorphic encryption with an inner product protocol to protect user privacy during face recognition. Their system achieves 98.78\% accuracy for both plaintext and ciphertext face recognition, demonstrating that encryption does not compromise performance. While this work provides a robust solution for privacy-preserving face recognition, we extend their approach by using fully homomorphic encryption (FHE), which provides enhanced flexibility and computational power compared to Pallier encryption. In our study, we conduct experiments using 128d, 512d, and 4096d embeddings, offering a more detailed analysis of how embedding dimensions impact recognition accuracy and efficiency. Unlike their approach, which primarily focuses on a fixed encryption protocol, we provide a complete Python-based framework that simplifies the integration of secure face recognition into real-world applications. This makes our system not only more versatile but also easier to implement and adopt by users, thus making privacy-preserving face recognition more accessible to the broader research and developer community.

The paper "Secure Face Matching Using Fully Homomorphic Encryption" \cite{related1} focuses on utilizing fully homomorphic encryption (FHE) for secure face recognition, allowing template matching to be performed directly on encrypted data. This work demonstrates the feasibility of secure face matching by reducing face templates to a 16 KB size, achieving a matching time of 0.01 seconds per face. They primarily use cosine distance for matching and conduct experiments with 128d and 512d embeddings. In contrast, our study not only adopts FHE for secure face recognition but extends it by testing with embeddings of 128d, 512d, and 4096d, which offers a more comprehensive evaluation of how different embedding dimensions affect performance. We also employ Euclidean distance for face matching, which provides a competitive alternative distance metric compared to their use of cosine distance. Most notably, our work offers a Python-based framework, which allows users to easily implement secure face recognition. This user-friendly framework is designed to facilitate the adoption of privacy-preserving face recognition systems, making our approach more accessible and practical for a broader audience.

\section{A Facial Recognition Pipeline}

\begin{figure}[H]
    \centering
    \includegraphics[width=0.95\textwidth]{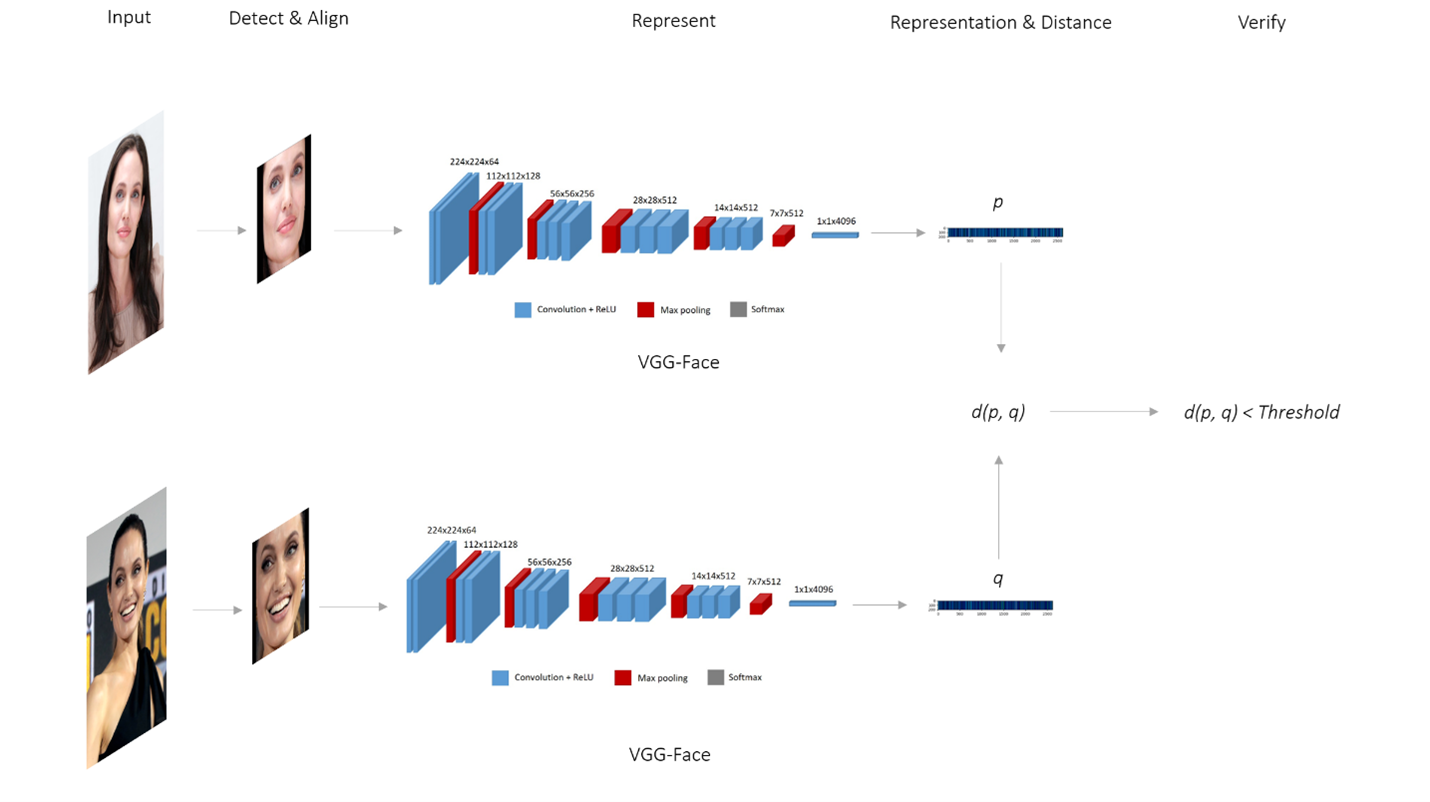}
    \caption{Facial Recognition Pipeline}
    \label{fig:pipeline}
\end{figure}

A modern facial recognition pipeline generally consists of four main stages: detection, alignment, representation, and verification \cite{fbdeepface} as illustrated in Figure \ref{fig:pipeline}. Each stage plays a critical role in ensuring the accuracy and reliability of the recognition process. A recent research indicates that the detection stage significantly influences accuracy, contributing up to 42\% of the overall performance, while alignment adds another 6\% \cite{deepface}. These findings highlight the importance of accurately identifying and preparing facial regions before representation and verification.

The representation stage, dominated by convolutional neural network (CNN) models, is at the core of modern facial recognition systems. CNN-based models have achieved remarkable accuracy, surpassing the 97.5\% human-level performance \cite{humanlevelaccuracy} on datasets such as the Labeled Faces in the Wild (LFW) \cite{lfw}. These models generate multidimensional vector embeddings that encapsulate the unique features of a face. The distance between two embeddings is then calculated to compare the similarity between faces. It is generally assumed that embeddings of the same person will have smaller distances, while those of different individuals will have larger ones.

Euclidean distance and cosine distance are two widely used metrics for comparing vector embeddings. The calculated distance is compared against a pre-tuned threshold to make a classification. If the distance is below the threshold, the pair is classified as belonging to the same person; otherwise, they are classified as different individuals.

Among the four stages, representation is the most computationally expensive step, requiring significant processing power to generate accurate embeddings as shown in Table \ref{tab:costs}. In contrast, the verification stage—calculating the distance between pre-computed embeddings and making classifications—is relatively lightweight. This balance allows modern facial recognition systems to be efficient, particularly when embeddings are pre-calculated and stored securely.

\begin{table}[H]
\centering
\caption{Plain Performances of Various Facial Recognition Models}
\begin{tabular}{lllcc}
\hline
Model       & Embedding Dims & Detect + Align + Represent & Verify \\
\hline
FaceNet128d & 128            & 1.59s/it       & 1333.04it/s  \\
FaceNet512d & 512            & 1.65s/it       & 1569.43it/s  \\
VGG-Face    & 4096           & 1.21s/it       & 812.75it/s   \\
\hline
\end{tabular}
\label{tab:costs}
\end{table}

\section{Homomorphic Encryption}

Homomorphic encryption allows computations to be performed on encrypted data, such that the result, when decrypted, is the same as if the operation had been performed on the plaintext data. This property is particularly useful for scenarios where data privacy must be maintained while still allowing computations to be done on the data, such as in cloud-based systems. 

In additive homomorphism, the encryption scheme supports the addition of ciphertexts. For example, in the Paillier encryption scheme, the ciphertexts can be added together to produce a result that corresponds to the sum of the plaintexts:

\begin{equation}
\label{eq:homomorphic_computation}
    {D(E(A) + E(B)) = A + B}
\end{equation}

where \( E \) represents encryption and \( D \) stands for decryption.

In multiplicative homomorphism, the encryption scheme supports the multiplication of ciphertexts. For example, the RSA and ElGamal encryption schemes allow ciphertexts to be multiplied together to produce a result corresponding to the product of the plaintexts:

\begin{equation}
\label{eq:homomorphic_computation_multiply}
    {D(E(A) \cdot E(B)) = A \times B}
\end{equation}

If an algorithm shows additively or multiplicatively homomorphic features, then it is called partially homomorphic (PHE). On the other hand, a fully homomorphic encryption (FHE) scheme supports both additive and multiplicative homomorphisms. This means it allows complex operations, such as both addition and multiplication, to be performed on encrypted data, enabling more flexible computations while maintaining the confidentiality of the data. Fully Homomorphic Encryption enables the execution of secure computations without ever needing to decrypt the data.

\section{Novel Encrypted Distance Computation in Homomorphic Encryption}

The Euclidean distance between two n-dimensional vectors \( \mathbf{\alpha} \) and \( \mathbf{\beta} \) is given by the Formula \ref{eq:euclidean} whereas the Cosine distance between two vectors \( \mathbf{\alpha} \) and \( \mathbf{\beta} \) is given by the Formula \ref{eq:cosine}. However, fully homomorphic encryption (FHE) has inherent limitations, making it impossible to compute these distances directly in their original forms. To address these challenges, we can apply specific workarounds to compute distances on encrypted data effectively.

\begin{equation}
\label{eq:euclidean}
    {d(\mathbf{\alpha}, \mathbf{\beta}) = \sqrt{\sum_{i=1}^{n} (\alpha_{i} - \beta_{i})^2}}
\end{equation}

\begin{equation}
\label{eq:cosine}
    {\theta(\mathbf{\alpha}, \mathbf{\beta}) = 1 - \frac{\mathbf{\alpha} \cdot \mathbf{\beta}}{\|\mathbf{\alpha}\| \|\mathbf{\beta}\|}} 
\end{equation}

The computation of Euclidean distance requires a sequence of operations: element-wise subtraction, dot products, and square root. While all these operations can be defined within the framework of fully homomorphic encryption, the square root calculation poses a challenge. To address this, we developed a workaround where the cloud computes the encrypted squared Euclidean distance, and on the on-premise side, we compare the decrypted value to a pre-tuned squared threshold as shown in Formula \ref{eq:euclidean_v2}. This method allows verification with Euclidean distance without directly requiring the square root calculation in the encrypted domain.

\begin{equation}
\label{eq:euclidean_v2}
    {d(\mathbf{\alpha}, \mathbf{\beta})^2 = \sum_{i=1}^{n} (\alpha_{i} - \beta_{i})^2}
\end{equation}

Encrypted cosine distance, however, presents more significant challenges. In Euclidean distance, the squared value of each dimension ensures that the result is always positive, even when the value is negative. This is not the case for cosine similarity, where negative values in the dot product can lead to issues. Specifically, when the dot product in cosine similarity is negative, encryption and decryption can result in the loss of the original value. To mitigate this, we performed min-max normalization on the embeddings before encryption, ensuring that all values remain within a defined positive range. Additionally, cosine similarity involves dividing the dot product by the norms of each vector, which creates further complications. Division is not supported in homomorphic encryption, and attempting to move the inverse of the norms into the numerator could cause an out-of-bound error in the scale.

To overcome this, we normalize each vector by dividing it by its magnitude on-premises before encryption as shown in Formula \ref{eq:cosine_updated} where $\hat{\alpha}$ and $\hat{\beta}$ are normalized vectors as illustrated in Formula \ref{eq:cosine_updated_v2} and Formula \ref{eq:cosine_updated_v3}. These normalized vectors are then encrypted and stored in the cloud. When calculating the cosine similarity, the dot product of the two encrypted normalized vectors is computed in the cloud. To convert the similarity into a cosine distance, the result is subtracted from 1. To achieve this, a tensor containing only the value 1 is created, encrypted with the public key, and then used in a subtraction operation with the encrypted similarity. This ensures the calculation of encrypted cosine distance entirely within the cloud.

\begin{equation}
\label{eq:cosine_updated}
    {\theta(\mathbf{\alpha}, \mathbf{\beta}) = 1 - \mathbf{\hat{\alpha}} \cdot \mathbf{\hat{\beta}} } 
\end{equation}

\begin{equation}
\label{eq:cosine_updated_v2}
    { \mathbf{\hat{\alpha}} = \frac{\alpha}{\|\mathbf{\alpha}\|}} 
\end{equation}

\begin{equation}
\label{eq:cosine_updated_v3}
    { \mathbf{\hat{\beta}} = \frac{\beta}{\|\mathbf{\beta}\|}} 
\end{equation}

On the on-premises side, the cosine distance is obtained by decrypting the result. It is then compared to a pre-tuned threshold to make the final decision. This updated approach streamlines the process, avoids out-of-bound errors, and ensures the homomorphic computation of cosine distance remains both feasible and accurate.

\begin{algorithm}[H]
\caption{Compute Embeddings, Encrypt Them On-Premises, and Store Them in the Cloud}
\label{alg:compute_embeddings}
\begin{algorithmic}[1]
\Require List of plain facial images $faces$, public key $public\_key$, distance metric $distance\_metric$
\Ensure Encrypted embeddings stored in $encrypted\_embeddings$. If distance metric is cosine, also store the norm

\State $encrypted\_embeddings \gets []$
\For{$i \gets 0$ \textbf{to} $|faces| - 1$}
    \State $current\_face \gets faces[i]$
    \State $embedding \gets \text{DeepFace.represent}(current\_face)$
    \If{$distance\_metric = \text{"cosine"}$} \Comment{Apply Min-Max Normalization to ensure all values are positive}
        \State $embedding \gets \frac{embedding - global\_min}{global\_max - global\_min}$
		\State $norm\_embedding \gets \text{norm}(embedding)$
		\State $embedding \gets \frac{embedding}{norm\_embedding}$
    \EndIf
    \State $encrypted\_embedding \gets \text{encrypt}(embedding, public\_key)$
	\State $encrypted\_embeddings.append(encrypted\_embedding)$
\EndFor
\State \Return $encrypted\_embeddings$
\end{algorithmic}
\end{algorithm}


\begin{algorithm}[H]
\caption{Compute Encrypted Euclidean Distances or Cosine Similarities in the Cloud}
\label{alg:homomorphic_computations}
\begin{algorithmic}[1]
\Require Target image $target\_img$, list of encrypted embeddings $encrypted\_embeddings$, distance metric $distance\_metric$, public key $public\_key$
\Ensure List of results with either encrypted squared Euclidean distances or cosine similarity

\State $target\_embedding \gets \text{DeepFace.represent}(target\_img)$

\If{$distance\_metric = \text{"cosine"}$}
    \State $norm\_target\_embedding \gets \text{norm}(target\_embedding)$
    \State $target\_embedding \gets \frac{target\_embedding}{norm\_target\_embedding}$
\EndIf

\State $encrypted\_target\_embedding \gets \text{encrypt}(target\_embedding, public\_key)$

\State $one = \text{encrypt}(1, public\_key)$
\State $results \gets []$
\For{$i \gets 0$ \textbf{to} $|encrypted\_embeddings| - 1$}
	\State $encrypted\_source\_embedding \gets encrypted\_embeddings[i]$
    \If{$distance\_metric = \text{"euclidean"}$}
        \State $difference \gets encrypted\_source\_embedding - encrypted\_target\_embedding$
        \State $distance \gets difference \cdot difference$ \Comment{Dot product to square the difference}
    \ElsIf{$distance\_metric = \text{"cosine"}$}
        \State $distance \gets one - encrypted\_source\_embedding \cdot encrypted\_target\_embedding$
    \EndIf
	\State $results.append(distance)$
\EndFor

\State \Return $results$
\end{algorithmic}
\end{algorithm}


\begin{algorithm}[H]
\caption{Perform Facial Recognition Using Encrypted Results On-Premises}
\label{alg:face_recognition}
\begin{algorithmic}[1]
\Require List of encrypted distances $encrypted\_results$, distance metric $distance\_metric$, private key $private\_key$, threshold $threshold$
\Ensure Identify if the target image is found in the facial database

\For{$i \gets 0$ \textbf{to} $|encrypted\_results| - 1$}
	\State $encrypted\_distance \gets encrypted\_results[i]$
	\State $distance \gets \text{decrypt}(encrypted\_distance, private\_key)$
	
    \If{$distance\_metric = \text{"euclidean"}$}    
        \If{$distance < threshold \times threshold$}
            \State \textbf{Output:} \texttt{"Target image is found in $i$-th item of facial database"}
            \State \textbf{break}
        \EndIf
    \ElsIf{$distance\_metric = \text{"cosine"}$}
        \If{$distance < threshold$}
            \State \textbf{Output:} \texttt{"Target image is found in $i$-th item of facial database"}
            \State \textbf{break}
        \EndIf
    \EndIf
\EndFor
\end{algorithmic}
\end{algorithm}


In summary, the generation of embeddings for our facial database is performed on the on-premises side, which holds both private and public keys. The public key is used to encrypt the embeddings of our facial database. If cosine distance is used as the distance metric, the on-premises system also normalizes the embeddings. These encrypted embeddings are then stored in the cloud. This process is described in Algorithm \ref{alg:compute_embeddings}.

On the cloud side, only the public key is available. When searching for an identity, the embedding of the target image is first calculated. This can be done in the cloud or even on an edge device. Similarly, if cosine distance is used, the cloud or edge device should normalize the target embedding. The target embedding is then encrypted using the public key. The cloud will compute the encrypted squared Euclidean distance or encrypted cosine distance for each comparison between the target image and the items in our facial database. These steps are detailed in Algorithm \ref{alg:homomorphic_computations}.

Finally, once the on-premises system has the list of encrypted distances, it will decrypt them one by one. If Euclidean distance is used, it compares the plain distance with the squared threshold value. If cosine is used, the distance is compared with the threshold itself. If the comparison is true, the identity is considered found. This process is explained in Algorithm \ref{alg:face_recognition}.

\section{Experiments}

The experiments were conducted to evaluate the feasibility of using homomorphic encryption in facial recognition systems. Two cryptosystem configurations were tested: a smaller configuration with moderate encryption parameters and a larger configuration with more robust encryption settings as shown in Table \ref{tab:first_experiment}. In the table, n represents the polynomial modulus degree, q denotes the bit sizes of the modulus for the coefficients, and g stands for the global scale.

Both configurations offer a 128-bit security level, as indicated by SEAL’s manual \cite{seal}, providing a strong encryption foundation. This configuration is considered secure beyond 2030, ensuring long-term protection against potential threats. These configurations were assessed using two distance metrics, Euclidean and Cosine, across three embedding dimensions to account for varying levels of data complexity. The Labeled Faces in the Wild (LFW) dataset’s test set, consisting of 1,000 image pairs evenly split between same-person and different-person labels, was used to benchmark the system. Performance was measured in terms of encryption, decryption, and homomorphic computation times. Homomorphic computation refers to the ability to perform distance calculations (such as Euclidean or Cosine distances) directly on encrypted data without decrypting it, ensuring privacy. 

\begin{table}[H]
\centering
\caption{Encrypted Facial Recognition Experiments}
\begin{tabular}{llll|l|ccc}
\hline
& &  & & &  &  Performance (ms / it) & \\
   & &  & & & & & \\
n & q & g & Distance & Operation & FaceNet & FaceNet & VGG-Face \\
  & & & & & (128d) & (512d) & (4096d)  \\
    & & & & & & & \\
\hline
     &     & &           & Encryption                &  7.19 & 7.40 & 7.56 \\
$2^{13}$ & 200 & $2^{40}$ & Euclidean & Homomorphic &  29.18 & 34.56 & 42.66 \\
     &     & &           & Decryption                &  1.85 & 1.97 & 1.99 \\
\hline
     &     & &           & Encryption                &  	7.72  & 8.34  & 8.06 \\
$2^{13}$ & 200 & $2^{40}$ & Cosine    & Homomorphic  &   29.36 & 35.27 & 43.88 \\
     &     & &           & Decryption                &  	1.99  & 2.15  & 1.88 \\
\hline
      &     & &           & Encryption & 25.63 & 24.14 & 26.13 \\
$2^{14}$ & 422 & $2^{60}$ & Euclidean & Homomorphic & 239.81 & 278.55 & 392.16 \\
      &     & &           & Decryption & 15.21 & 14.12 & 16.30 \\
\hline
      &     & &           & Encryption              & 24.39  & 24.10  & 24.47 \\
$2^{14}$ & 422 & $2^{60}$ & Cosine    & Homomorphic & 236.64 & 308.64 & 347.22 \\
      &     & &           & Decryption              & 13.11  & 12.95  & 13.75 \\
\hline
\end{tabular}
\label{tab:first_experiment}
\end{table}

In the smaller configuration, operations such as encryption and decryption demonstrated low latency, making this setup suitable for applications prioritizing efficiency. Homomorphic computations exhibited an increase in time with larger embedding dimensions, as expected due to the higher complexity of the data. However, the processing times remained within an acceptable range for real-time or near-real-time applications. Between the two distance metrics, Cosine computations were slightly more resource-intensive compared to Euclidean computations.

The larger configuration introduced a notable increase in computational overhead. Encryption and decryption times were higher compared to the smaller configuration, reflecting the enhanced security parameters. Similarly, homomorphic computations took significantly longer, especially for embeddings with higher dimensions. The added computational cost of this configuration underscores the trade-off between stronger encryption and processing efficiency. Nevertheless, the system maintained reliable performance across all scenarios, demonstrating its ability to handle more secure encryption settings while still achieving accurate results.

Overall, the experiments revealed that embedding dimensions play a significant role in determining processing times across all operations. Larger dimensions resulted in longer processing times due to the increased data size, but the system handled these variations effectively. Additionally, while the larger configuration incurred greater computational costs, it provided enhanced encryption strength, making it ideal for scenarios requiring heightened security. This underscores the robustness of the proposed approach, demonstrating that homomorphic encryption can be seamlessly integrated into facial recognition systems without compromising accuracy. The study highlights a balanced approach to achieving secure and efficient recognition, paving the way for future optimizations and broader real-world adoption.

In summary, adding homomorphic encryption introduces computational overhead for encryption, homomorphic distance computation, and decryption, yet these operations occur in milliseconds, making this approach feasible for secure facial recognition. For example, with FaceNet-128d, the detect + align + represent phase takes 1.59 s/it, and verification without encryption takes 0.75 ms/it. When homomorphic encryption is added (7.19 ms/it for encryption, 29.18 ms/it for homomorphic computation, and 1.85 ms/it for decryption), the total increases to 38.07 ms/it, resulting in a total of 1628.82 ms/it. This represents a 2.4\% increase in total processing time compared to the plain process, which is still negligible when compared to the 1590.75 ms/it required for the embedding computation. Similarly, for FaceNet-512d, the total time increases from 1665.75 ms/it to 1705.68 ms/it, representing a 2.4\% additional cost, and for VGG-Face, the total time increases from 1212.75 ms/it to 1257.96 ms/it, representing a 3.7\% additional cost.

For the larger configuration with FaceNet-128d, the total time increases from 1590.75 ms/it to 1875.56 ms/it, representing an 18\% additional cost. For FaceNet-512d, the total time increases from 1665.75 ms/it to 1959.74 ms/it, representing a 17.6\% additional cost. For VGG-Face, the total time increases from 1212.75 ms/it to 1661.74 ms/it, representing a 37\% additional cost.

\subsection{Environment}

The experiments in this study were conducted on a machine running a Linux environment within the Windows Subsystem for Linux (WSL 2) framework. The system used the Linux kernel version 5.15.167.4-microsoft-standard-WSL2 on an x86-64 architecture. It featured an 11th Generation Intel(R) Core(TM) i7-11370H CPU, operating at a base frequency of 3.30 GHz, with 4 cores and 8 threads, supporting virtualization via VT-x. The machine was equipped with 15 GB of RAM, with a swap space of 4 GB, and had sufficient disk storage, with 953 GB total disk space. The computational environment was further enhanced by leveraging the flexibility and compatibility of WSL 2, providing a robust platform for conducting secure and efficient experiments.

\section{Conclusion}

In this work, we introduce CipherFace, a first-of-its-kind algorithm that enables high-performance homomorphic encryption-based facial recognition through novel computational transformations. CipherFace fundamentally changes how encrypted embeddings are manipulated, demonstrating that complex recognition operations can be executed in encrypted domains without the traditional computational overhead. Through careful algorithmic design choices, we tackle the fundamental computational challenges that have historically prevented practical deployment of homomorphic encryption in real-world facial recognition systems. Our method enables cloud-side processing of encrypted embeddings with remarkably reduced computational costs compared to existing approaches, while maintaining robust security throughout the recognition pipeline.

We validate our algorithmic contributions through comprehensive experimental evaluation. We present a new encrypted distance computation protocol that efficiently handles both Euclidean and Cosine distance calculations while preserving security guarantees. Extensive testing on the Labeled Faces in the Wild dataset reveals that our approach delivers significant performance improvements across various embedding dimensions from widely-adopted models including FaceNet128, FaceNet512, and VGG-Face. By implementing NIST-recommended 128-bit security configurations, the system ensures long-term protection of sensitive biometric data. These results demonstrate that our approach successfully bridges the gap between theoretical homomorphic encryption capabilities and the practical demands of facial recognition applications.

The framework simplifies client-side operations by offloading resource-intensive computations, such as distance calculations, to the cloud. This eliminates the need for additional on-premises processing, reducing computational overhead while retaining control over sensitive data. Moreover, the approach scales well with increasing database sizes and embedding dimensionalities, making it a practical solution for real-world facial recognition applications.

In summary, CipherFace combines the strengths of homomorphic encryption with modern facial recognition techniques to provide a secure, efficient, and scalable solution for identity verification and recognition in cloud-based environments. This work paves the way for more secure implementations of machine learning models in sensitive applications, balancing privacy and performance. Future research could explore optimizing the encryption schemes, enhancing the encrypted distance computation methods, and extending the framework to support broader use cases in privacy-preserving artificial intelligence.

\newpage

\bibliographystyle{unsrt}

\end{document}